\documentclass[aps,twocolumn,floats,floatfix,showpacs,superscriptaddress]{revtex4}
\usepackage{epsfig}
\begin{document}

\title{Dynamic Fracture Model for Acoustic Emission}
\author{Manuela Minozzi}
\affiliation{INFM UdR Roma 1 - Dipartimento di Fisica, 
Universit\`a "La Sapienza", P.le A. Moro 2,
00185 - Roma, Italy}
\affiliation{INFM UdR Roma 3, Dipartimento di Fisica,
Universit\`a di Roma 3, via della vasca navale 84, 00146
Roma, Italy.}
\affiliation{CNR Istituto di Acustica ``O. M. Corbino'', via 
del fosso del Cavaliere 100, 00133 Roma, Italy}
\author{Guido Caldarelli}
\affiliation{INFM UdR Roma 1 - Dipartimento di Fisica, 
Universit\`a "La Sapienza", P.le A. Moro 2,
00185 - Roma, Italy}
\author{Luciano Pietronero}
\affiliation{INFM UdR Roma 1 - Dipartimento di Fisica, 
Universit\`a "La Sapienza", P.le A. Moro 2,
00185 - Roma, Italy}
\affiliation{CNR Istituto di Acustica ``O. M. Corbino'', via 
del fosso del Cavaliere 100, 00133 Roma, Italy}
\author{Stefano Zapperi}
\affiliation{INFM UdR Roma 1 - Dipartimento di Fisica, 
Universit\`a "La Sapienza", P.le A. Moro 2,
00185 - Roma, Italy}

\date{\today}

\begin{abstract}
We study the acoustic emission produced by micro-cracks using a two-dimensional
disordered lattice model of dynamic fracture, which allows to relate the acoustic response to the internal damage of
the sample. We find that the distributions of acoustic 
energy bursts decays as a power law in agreement with 
experimental observations. The scaling exponents measured in
the present dynamic model can related to those obtained in the
quasi-static random fuse model. 
\end{abstract}
\maketitle

Crackling noise \cite{SET-01} is widely observed in systems as different as
superconductors \cite{flux}, magnets \cite{DUR-01}
or  plastically deforming crystals \cite{MIG-01}.
A typical example is the acoustic emission (AE) recorded
in a stressed material before failure. The noise is a consequence
of micro-cracks forming and propagating in the material and should
thus provide an indirect measure of the damage accumulated in
the system. For this reason, AE is often used as a non-destructive
tool in material testing and evaluation.
Beside these practical applications, understanding the statistical
properties of crackling noise has become a challenging theoretical
problem. The distribution of crackle amplitudes follows a power law,
suggesting an interpretation in terms of critical phenomena and
scaling theories. This behavior has been observed in several
materials such as wood \cite{ciliberto}, cellular glass
\cite{strauven}, concrete \cite{ae} and paper \cite{paper} 
to name just a few. 

The statistical properties of fracture in disordered media 
are captured qualitatively  by lattice models,  describing the medium as 
a discrete set of elastic bonds with randomly 
distributed failure thresholds \cite{hr,fuse,duxbury}. 
After each crack the stress is redistributed in the lattice 
in a quasi-static approximation: 
i.e. the crack velocity is much slower than stress relaxation.
Thus acoustic waves are not taken into account and the activity is
monitored by the damage evolution or by the dissipated elastic energy.
Numerical simulations indicate that micro-cracks propagate in avalanches 
giving rise to an heterogeneous response. The avalanche distribution
is typically described by power law distributions and
the results are usually interpreted in the framework of phase
transitions \cite{hansen,zrvs,zvs,gcalda,alava}. 
Despite the fact that critical phenomena are normally
associated with a certain degree of universality (i.e. the scaling
exponents should not depend on micro-structural details), there
has been so far no quantitative agreement between models and experiments.
A reason that could account for this discrepancy is the absence
of acoustic waves in most models.  It is then not obvious how to relate
AE activity to internal avalanches.

Dynamic lattice models have been widely used in the past to
analyze fracture processes \cite{dyn,marder,RAU,mar}, but although
acoustic waves are explicitly included, the AE signal is usually
not analyzed. Here we use a lattice model for dynamic fracture 
in a disordered medium, to obtain a direct correspondence between 
the recorded AE activity and the internal damage evolution. 
We find that the cumulative AE amplitudes
are directly related --- by a power law --- to the cumulative damage. 
Next, we measure the distribution of the AE burst energies and
find a power law with an exponent $\beta \simeq 1.7$ independent on 
the loading rate. This exponent can be related to the exponent
describing failure avalanches in quasi-static models
\cite{hansen,zrvs,zvs,gcalda,alava}.

We consider a scalar model of dynamic fractures where a 
two-dimensional lattice is loaded in mode III \cite{marder}:
the lattice lies in the $(x,y)$ plane and deformation occurs
along the $z$ axis, so that the equations of elasticity
become scalar. The equation of the motion for the anti-planar 
displacement $u$ of a site with coordinate $i,j$ is 
\begin{equation}
\rho\ddot{u}_{i,j}= -K\sum_{(l,m)}  (u_{i,j}-u_{l,m})-\Gamma\dot{u}_{i,j},
\label{eq:1}
\end{equation}
where the sum runs over the nearest neighbors $(l,m)$ of site $(i,j)$, 
$K$ is the elastic constant, $\rho$ is the density
and dissipation is simulated by a viscous damping with a constant $\Gamma$. 
In order to suppress some lattice effects, we use  a 45 degree tilted 
square lattice. A constant strain rate is imposed to the model,
by moving the boundary sites on two opposite boundaries at constant
velocity $V$ and $-V$, respectively. Periodic boundary conditions
are imposed in the other direction. Disorder is simulated assigning 
randomly distributed failure threshold: a bond is removed (i.e. $K$ 
is set to zero) when $\Delta u > f_c$, and $f_c$ is uniformly distributed
in $[0,1]$. Notice that in the quasi-static limit 
($V\to 0$, $\rho\to 0$, $\Gamma\to 0$)  the model reduces 
to the random fuse model (RFM), where a lattice of fuses with
random threshold are subject to an increasing voltage \cite{fuse,duxbury}. 
Due to the scalar nature of our model there is a direct mapping
between elastic and electric parameters \cite{hr}. 

The equation of motion (Eq.~\ref{eq:1}) is integrated numerically
using a fourth order Runge-Kutta method. We work with a lattice
of linear size $L=80$ and chose the units of space and time
so that $\rho=K=1$. Each time a bond is stretched
beyond its threshold the lattice constant is set to zero and
an elastic wave is emitted. Due to the anti-plane constraint
for the displacements, we only have transverse wave propagation
with sound speed $c=\sqrt{K/\rho}=1$ in our units. The damping
constant is chosen to be $\Gamma=0.1$ so that typical length traveled
by a wave is a little smaller than the lattice size. For smaller
values of $\Gamma$ ringing effects and reflected waves do not
allow to separate the single pulses and the lattice breaks at
once. On the other hand, excessive damping leads to very
small acoustic activity and the sample breaks suddenly at the edges.
Even if the damping constant is small reflected waves
can induce boundary failure, due to the rigidity of the 
loaded edge. Thus we do not allow that bonds fail in two boundary layers
of length $l=5$ close to the loaded edges. This corresponds to
apply a load through a soft contact. The model is simulated
for a variety of loading velocities all much lower than the
sound speed $V \ll c$. 

\begin{figure}[t]
\centerline{\psfig{file=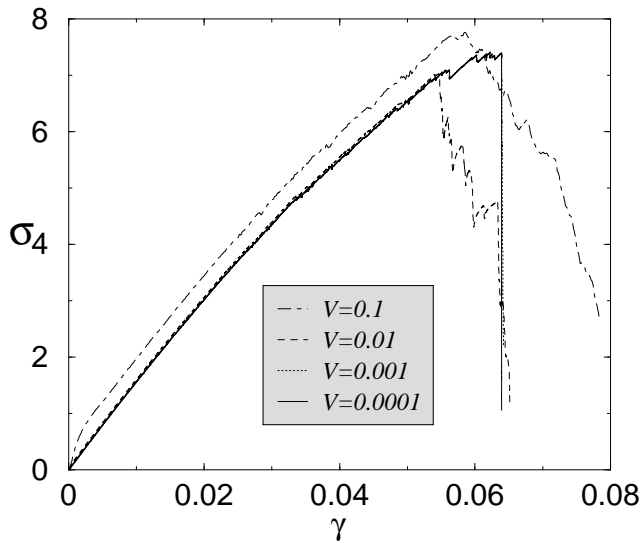,width=8.5cm,clip=!}}
\caption{The stress-strain curve for different applied strain rates. }
\label{fig:1}
\end{figure}

Measuring the displacements of every lattice site and calculating
the forces for every time steps, we have obtained
 the stress-strain curve for four different value of 
the applied strain rate. In Fig.~\ref{fig:1} we show 
that the stress is a linear function of the strain up 
to the yield point, which precedes the total failure of the sample.
The applied strain rate has little effect on the linear part of the curve,
while it influences the curve after the yield point.
 
Monitoring the activity of some
particular lattice sites we have direct access to the AE
signal. These sites mimic the effect of transducers
coupled to the material in a typical AE experiment. 
In a typical run, we record the displacements, velocities and accelerations 
of four sites in the boundary layer and two sites in the interior. 
Typically, AE distributions are recorded from a single site
and averaged over ten realizations of the disorder.
We have tested that the statistical properties of the signal
do not vary for different boundary sites, while there is
a clear difference between boundary and inner sites.
In the following, we concentrate on sites in the boundary layer, 
in order to avoid excessive fluctuations due to failures 
occurring on neighboring bonds in the inner region.

\begin{figure}[h]
\centerline{\psfig{file=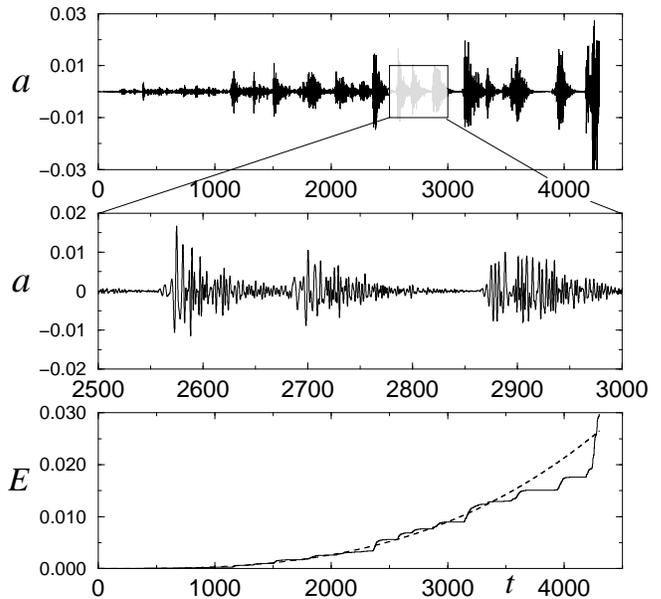,width=8.5cm,clip=!}}
\caption{A typical signal measured in the model. The upper panel 
shows the velocity acceleration of a site close to the boundary. 
In the middle panel we show a magnification of a portion of the signal. 
In the lower panel is reported the cumulative energy,
 $E(t)\equiv \int_0^t dt' a^2(t')$, as a function of time for a single
realization of disorder. The dashed line follows $t^3$. }
\label{fig:2}
\end{figure}

An example of the typical signals recorded with our model are
reported in Fig.~\ref{fig:2}. A large acoustic activity is visible in the upper panel where we show the 
local acceleration $a$ of a boundary site as a function of time. 
We can also monitor the velocity signal which is 
simply related to the acceleration and display the same features. In the present model,
it will be convenient to use the acceleration as a AE monitoring
tool, since the velocity has a bias induced by the external loading:
even in the absence of cracking the lattice has a non-vanishing velocity.
We define the associated cumulative energy as 
\begin{equation}
E(t)\equiv \int_0^t dt' a^2(t').
\end{equation}
The behavior of the cumulative acoustic energy $E(t)$ is typically monitored
in AE experiments. In some cases, $E(t)$ is found to increase as a power law 
\cite{ciliberto},
or exponentially in other cases \cite{paper}. In general one expects a
marked peak close to failure, as we also observe in Fig~\ref{fig:2},
obtained for $V=10^{-3}$.
The curve is well fitted by cubic law, $E \sim t^3$.

\begin{figure}[t]
\centerline{\psfig{file=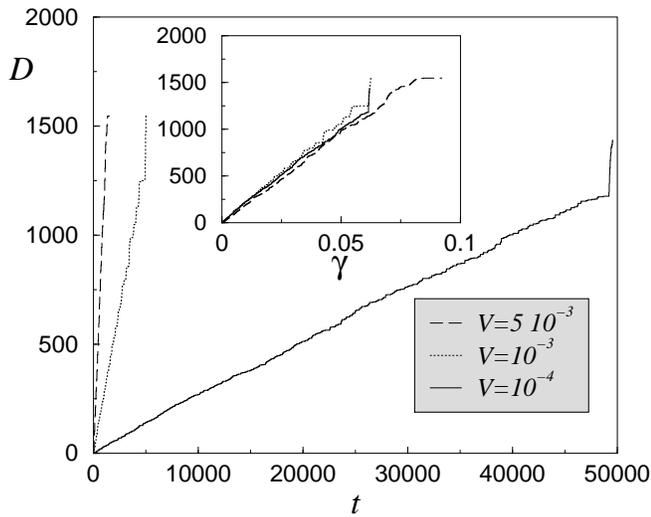,width=8.5cm,clip=!}}
\caption{The damage $D$ evolution is displayed as a function
of time for different loading rates. The curves collapse onto
each other when plotted with respect to strain $\gamma\equiv Vt/L$.}
\label{fig:3}
\end{figure}

\begin{figure}[t]
\centerline{\psfig{file=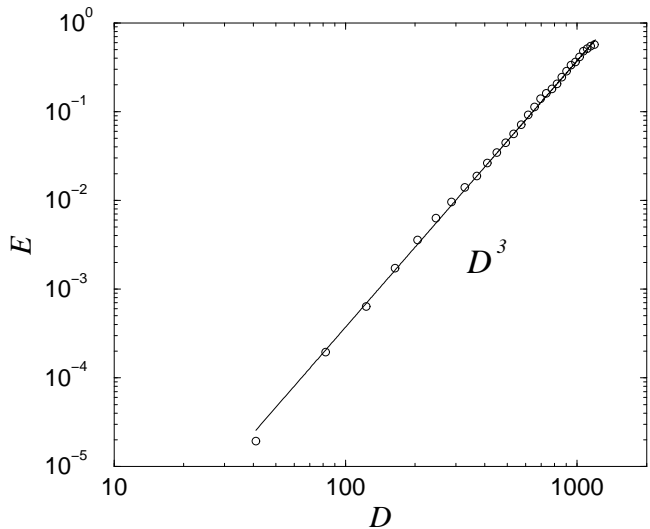,width=8.5cm,clip=!}}
\caption{The average cumulative energy plotted as a function
	of damage. The line represents a $D^3$ law which fits well the
	curve.}
\label{fig:4}
\end{figure}

A central problem in AE measurements is to correlate the recorded
acoustic activity with the internal damage state. In this way, AE
can be used as a tool for damage evaluation. In our model, we have
a direct access to the internal damage $D$ that can be defined
as the total number of failed bonds. We find that $D$ increases 
linearly with time (see Fig.~\ref{fig:3}) apart from a rapid
increase very close to failure. Rescaling the curves with the loading
rate one sees that $D$ is in fact a linear function
of the applied strain $\gamma\equiv (Vt)/L$ 
(see the inset of Fig.~\ref{fig:4}).

These observations thus lead to a direct scaling relation between
internal damage and released acoustic energy: Fig.~\ref{fig:4} 
shows that $E$ scales as expected as $D^3$. 
A direct consequence of this result is that 
the measured acoustic energy is proportional to the 
released elastic energy $E_{el}\sim KD\gamma^2 \sim D^3 \sim E$. 

A large amount of theoretical activity has been devoted in the past
to understand the origin of power law distributions of AE amplitudes
widely observed in material fracture. Most of the analysis was devoted
to quasi-static models, such as the the RFM, where
fracture was shown to occur in damage burst, distributed as
$P(D) \sim D^{-\tau}$ with $\tau \simeq 2.5$ \cite{hansen,zrvs}. 
This value is in perfect agreement with the result $\tau=5/2$ obtained 
exactly \cite{hansen1} for the exponent of the avalanche distribution of the 
fiber bundle model (FBM) \cite{dfbm}, 
where $N$ fibers with random failure threshold
are loaded in parallel. It was thus conjectured that the long-range
stress transfer present in the RFM was equivalent to the infinite
range load redistribution of the FBM, placing the two models into
the same universality class \cite{hansen,zrvs}. 
A similar exponent was found in a vectorial fracture model, 
so that this class could be even broader \cite{zrvs}. 
Comparing this result with AE experiments is problematic
since quasi-static models do not account for wave propagation. 

\begin{figure}[t]
\centerline{\psfig{file=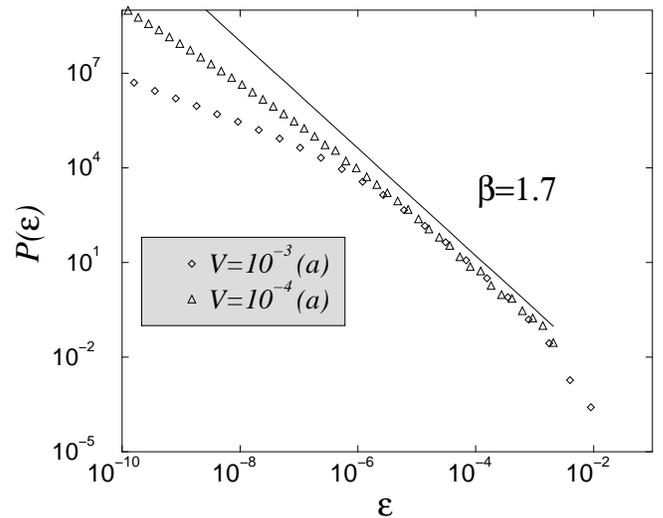,width=8.5cm,clip=!}}
\caption{The distribution of acoustic burst energies for different
	driving rates. The line is a power law with exponent $\beta=1.7$.}
\label{fig:5}
\end{figure}

Here we can directly measure the distribution of pulse sizes
due to the acoustic activity. In Fig.~\ref{fig:5} we report the
distribution of energies  for the acceleration 
signal, defining $\epsilon\equiv a^2$.
 In both cases the distribution
decays as power law with an exponent $\beta =1.7 \pm 0.1$,
independent on the loading rate, which only affects the 
low part of the distribution.
The same law is found in the case of the velocity signal.
Experimental results report an exponent value in the same range, 
even if it differs a little from one material to another:
for wood the exponent is $\beta =1.51 \pm 0.05$, for fiberglass $\beta =2.0 \pm 0.01$ \cite{ciliberto}, 
$\beta =1.30 \pm 0.1$ for paper \cite{paper},
 $\beta =1.5 \pm 0.1$ for experiments on cellular glass \cite{strauven}.

Using the scaling relation between released acoustic energy
and damage discussed above,  we can relate the exponent $\beta$ to $\tau$.
From $E \sim D^3$ and $D \sim t$, we expect $\epsilon \sim D^2$.
Substituting this expression in the equation for the probabilities
$P(\epsilon)d\epsilon=P(D)dD$, we obtain $\tau=1+2(\beta-1)=2.4$, which
is very close to $\tau=5/2$  measured in the RFM.
Thus we conjecture that the acoustic energy exponent measured 
in our dynamic model is directly
related with the damage exponent measured in the corresponding 
quasi-static model \cite{nota}.

In conclusions, we have introduced a lattice model
of dynamic fracture which can be used to model AE experiments.
The model allows to clarify important issues in the interpretation
of the experiments, namely the relation between internal damage
and released acoustic energy. In particular, we derive direct relations
between the scaling behavior of failure avalanches and acoustic bursts.
It would be interesting to generalize this analysis to more realistic
situations, exploring the role of dimensionality,
load conditions and lattice anisotropy.
However, in comparing the simulated signal with experiments, we should be 
careful  about the definition of the events in the time series, since 
the amplifier and the AE sensors could bias the recorded waveform, introducing a 
systematic error in the data.

This work has been supported by the European Network contract FMRXCT980183
and the INFM center SMC.


\begin{thebibliography}{99}
\bibitem{nota} The distribution of the ``elastic'' energy
	released in each avalanche was recently measured in
	the quasi-static FBM and found to decay as power law with 
	exponent $\beta\simeq 1.8$ \protect\cite{paper},
	 in agreement with our results.
\end{thebibliography}
\end{document}